\documentclass[a4paper,11pt]{article}
\usepackage{pos}

\usepackage{graphicx}
\usepackage{amsmath}
\usepackage{hyperref}
\usepackage{axodraw2}
\usepackage{overpic}
\usepackage[capitalize]{cleveref}
\usepackage{enumitem}

\newcommand{\meleven}{\cdot 10^{-11}}  

\title{Short-distance contributions to Hadronic-light-by-light for the muon $g-2$}
\author*[a]{Johan Bijnens}
\author[a,b]{Nils Hermansson-Truedsson}
\author[c]{Antonio~Rodr\'{i}guez-S\'{a}nchez}

\affiliation[a]{Division of Particle and Nuclear Physics, Department of Physics, Lund University,
Box 118, \\SE 221 00 Lund, Sweden}
\affiliation[b]{Higgs Centre for Theoretical Physics, School of Physics and Astronomy, The University of Edinburgh, James Clerk Maxwell Building,
Peter Guthrie Tait Road,
Edinburgh,
EH9 3FD}
\affiliation[c]{Instituto de Física Corpuscular, Universitat de València — CSIC, Parque Científico, Catedrático José Beltrán 2, E-46980 Paterna, Spain}

\emailAdd{johan.bijnens@fysik.lu.se}
\emailAdd{nils.hermansson-truedsson@ed.ac.uk}
\emailAdd{anrosanz@ific.uv.es}

\abstract{
  This talk discusses short-distance contributions to the hadronic light-by-light part of the muon g-2 as in the Standard Model. A short discussion about the theory prediction is followed by the status of our work. The main new results since the previous chiral dynamics workshop is how to calculate the higher order corrections to the Melnikov-Vainshtein region using the operator product expansion. We work out fully the next order in the OPE, $D=4$, including photon and gluon only operators.
  In addition we briefly discuss the ongoing work on gluonic corrections to the next-order and nonperturbative estimates.
  }

\FullConference{The 11th International Workshop on Chiral Dynamics (CD2024)\\
 26-30 August 2024\\
Ruhr University Bochum, Germany\\}

\begin{document}
\maketitle

\section{Introduction}

\begin{figure}
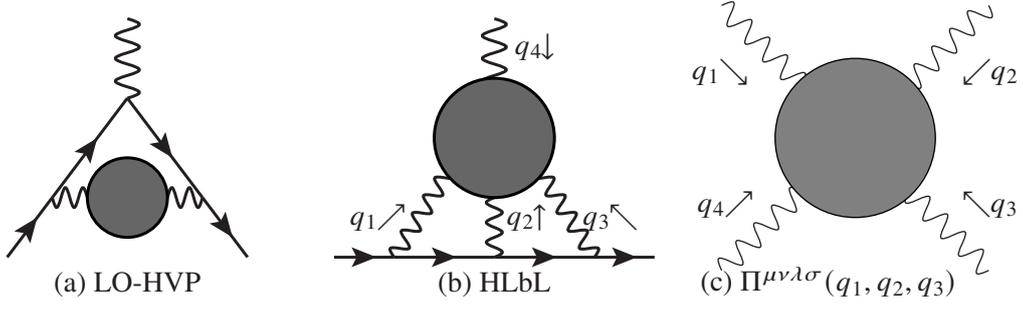

\begin{center}
\begin{axopicture}(60,80)(0,-10)
  \SetScale{1.5}
  \SetWidth{0.75}
  \Photon(30,60)(30,40){3}{3.5}
  \ArrowLine(0,0)(11.25,15)
  \ArrowLine(11.25,15)(30,40)
  \ArrowLine(30,40)(48.75,15)
  \ArrowLine(48.75,15)(60,0)
  \Photon(11.25,15)(20,15){-2.5}{2.}
  \Photon(40,15)(48.75,15){2.5}{2.}
  \GCirc(30,15){10}{0.4}
  \Text(30,-10)[b]{(a) LO-HVP}
  \end{axopicture}
  \hspace*{2cm}
  \setlength{\unitlength}{1pt}
  \begin{axopicture}(80,90)(0,-10)
  \SetScale{1.5}
  \SetWidth{0.75}
  \Photon(40,60)(40,40){3}{3.5}
  \ArrowLine(0,0)(15,0)
  \ArrowLine(15,0)(40,0)
  \ArrowLine(40,0)(65,0)
  \ArrowLine(65,0)(80,0)
  \Photon(15,0)(40,40){2}{9}
  \Photon(40,0)(40,40){-2}{8}
  \Photon(65,0)(40,40){-2}{9}
  \GCirc(40,30){15}{0.4}
  \Text(18,10)[r]{$q_1\!\!\nearrow$}
  \Text(43,10)[l]{$q_2\!\!\uparrow$}
  \Text(62,10)[l]{$q_3\!\!\nwarrow$}
  \Text(45,53)[l]{$q_4\!\!\downarrow$}
  \Text(40,-10)[b]{(b) HLbL}
  \end{axopicture}
  \hspace*{2cm}
  \raisebox{5pt}{
  \setlength{\unitlength}{1pt}
  \begin{axopicture}(100,100)
  \SetScale{1.0}
  %\SetWidth{1.0}
  \Photon(0,0)(100,100){4}{15}
  \Photon(0,100)(100,0){4}{15}
  \GCirc(50,50){30}{0.5}
  \Text(90,20)[lb]{$\nwarrow\!\!q_3$}
  \Text(90,80)[lt]{$\swarrow\!\!q_2$}
  \Text(10,80)[rt]{$q_1\!\!\searrow$}
  \Text(12,20)[rb]{$q_4\!\!\nearrow$}
  \Text(40,-10)[b]{(c) $\Pi^{\mu\nu\lambda\sigma} (q_1,q_2,q_3)$}
  \end{axopicture}
  }
\end{center}
\caption{(a) The hadronic vacuum polarization; (b) the hadronic light-by-light contribution; (c) the underlying four-point function.}
\label{fig:hadronic}
\end{figure}

The muon anomalous magnetic moment is one of the most precise measurements in particle physics with~\cite{Bennett:2006fi,Muong-2:2023cdq}
\begin{align}
  \label{eq:expresult}
a_\mu =(g_\mu-2)/{2}  = 0.00116592059(22)\text{ or a precision of } 1.9\cdot10^{-7}= 0.19\text{ ppm}.
\end{align}
The question is then, can this be calculated in the Standard Model to the same precision? The theory consensus was worked out in the white paper \cite{Aoyama:2020ynm} where an update will be pushed in the near future. The present status was discussed in the plenary talks of G.~Colangelo \cite{ColangeloPoS} and H.~Wittig \cite{WittigPoS}. Also relevant is the talk by S.~Holz \cite{HolzPoS}.
All theory uncertainties are under sufficient control except for the hadronic contributions where the two main ones are shown in \cref{fig:hadronic}.
The main uncertainty at present comes from the lowest-order (LO) hadronic vacuum polarization (HVP) where there is a discrepancy if calculated using experimental input \cite{ColangeloPoS} or lattice QCD \cite{WittigPoS}. The underlying hadronic object is a two-point of two electromagnetic currents. The main problem is that one needs a 0.13\% precision in order to match the precision of \cref{eq:expresult}. We will not discuss this contribution further.

The other main hadronic contribution, LO hadronic light-by-light (HLbL), has as underlying object a four-point function of electro-magnetic currents. This object is much more complicated than the two-point function, since it depends on six kinematic variables and has a large number of Lorentz-invariant functions involved. The amount of experimental data is also rather limited. The first proper calculations within models were done by two groups \cite{Hayakawa:1997rq,Hayakawa:2001bb} and \cite{Bijnens:1995cc,Bijnens:1995xf,Bijnens:2001cq}. Here it was found that the main contribution came from pseudo-scalar exchange which was studied by many authors e.g.~\cite{Knecht:2001qf,Melnikov:2003xd}. Ref.~\cite{Melnikov:2003xd} was also the first paper applying short-distance constraints to the full four-point function depicted in \cref{fig:hadronic}(c). A major problem was always how to separate in a consistent way estimates of different contributions.

The four-point function we need to discuss is $\Pi^{\mu\nu\lambda\sigma}$, which can be rewritten in terms of Lorentz invariant functions $\hat\Pi_i$ but what really contributes is \cite{Aldins:1970id}
\begin{align}
  \label{eq:derivpi}
  \left.\dfrac{\partial}{\partial q_{4\rho}}\Pi^{\mu\nu\lambda\sigma}\right|_{q_4=0}\,.
\end{align}
Once this derivative is known we can project on a set of 12 functions and obtain the contribution to $a_\mu$ \cite{Colangelo:2015ama}
\begin{align}\label{eq:amuint}
  a_{\mu}^{\textrm{HLbL}} 
  = 
  \frac{2\alpha ^{3}}{3\pi ^{2}} 
  & \int _{0}^{\infty} dQ_{1}\int_{0}^{\infty} dQ_{2} \int _{-1}^{1}d\tau \, \sqrt{1-\tau ^{2}}\, Q_{1}^{3}Q_{2}^{3}
  \sum _{n=1}^{12} T_{n}(Q_{1},Q_{2},\tau)\, \overline{\Pi}_{n}(Q_{1},Q_{2},\tau)\, .
  \end{align}
  Here the variables $Q_{i}$ are the magnitudes of the Euclidean squares of the momenta $q_i$, $Q_i^2 = -q_i^2$, and $\tau$ relates the three scales through $Q_3^2=Q_1^2+Q_2^2+2\tau Q_1Q_2$. The kernels $T_i $ are known functions given in Ref.~\cite{Colangelo:2015ama} and the $\overline\Pi_i$ can be determined from the $\hat\Pi_i$.

The white paper \cite{Aoyama:2020ynm} improved significantly on the earlier work thanks to two major steps forward. In a series of papers starting with \cite{Colangelo:2015ama} showed how different contributions to the four-point function can be separated in a well defined way using cuts in intermediate states. This solved the overlap problem between different models for different contributions by uniquely identifying intermediate states. These authors together with collaborators also formulated dispersion relations for the form-factors needed for the various intermediate states. The other main step forward was the realization that short-distance contributions could be calculated from perturbative QCD even in the $q_4\to0$ limit needed \cite{Bijnens:2019ghy}. These allowed to obtain HLbL with controlled errors\footnote{The HLbL numerical estimates have been quite compatible since about 2000 but the errors were often questioned.}.
The different contributions can be summarized as, \cite{Aoyama:2020ynm} and references therein:
\begin{itemize}[noitemsep,nolistsep]
  \item ``Long distance'': under good control
  \begin{itemize}[noitemsep,nolistsep]
  \item Dispersive method: Berne group of G.~Colangelo and collaborators
  \item $\pi^0$ (and $\eta,\eta^\prime$) pole: $ 93.8(4.0)\meleven$\hskip1.5cm  (Note \cite{Bijnens:2001cq} had 85(13))
  \item Pion and kaon box (pure): $ -16.4(2)\meleven$
  \item $\pi\pi$-rescattering (include scalars below 1 GeV): $ -8(1)\meleven$
  \end{itemize}
  \item Charm (beauty, top) loop: $ 3(1)\meleven$
  \item ``Short and medium distance'' { Main source of the error}
  \begin{itemize}[noitemsep,nolistsep]
    \item Scalars, tensors: $ -1(3)\meleven$
  \item Axial vector:  $6(6)\meleven$
  \item Short-distance: $ 15(10)\meleven$
  \end{itemize}
  \item $ a_\mu^\text{HLbL}=92(19)\meleven$
  \end{itemize}
So the dispersive method together with our short-distance results allowed to get the error down and trustable. The main part of the $19\meleven$ error came from the uncertainty in intermediate distance contribution and possible overlap with the short-distance QCD estimates. It should be noticed that lattice QCD has since the appearance of \cite{Aoyama:2020ynm} produced results that are quite compatible
\begin{itemize}[noitemsep,nolistsep]
  \item RBC/UKQCD 23 \cite{Blum:2023vlm} $124.7(14.9)\meleven$  
  \item Mainz 21/22 \cite{Chao:2021tvp} $109.6(15.9)\meleven$
  \item BMW \cite {Fodor:2024jyn} $125.5(13)\meleven$
\end{itemize}

What has happened since then on the phenomenological front? The dispersive method has been applied to $\eta,\eta^\prime$ as well as to axials-vectors and first steps towards tensor contributions have been taken, a complete list of references can be found in \cite{Hoferichter:2024bae}. There is also work from the holographic QCD which can be traced from \cite{Mager:2025pvz}.

\section{Short-distance introduction}
In the remainder we will discuss only parts relevant to short-distance contributions. As an indication of the difficulties to be expected \cref{tab:resonances} lists the states between 1 and 1.5 GeV and in addition there are also multi-particle states. The couplings to on-shell photons are known for some of these and for off-shell photons there is very little experimental information. We thus need more theoretical methods. The improved data will help with the dispersive approach for improving around 1 GeV and provide constraints.
\begin{table}
\begin{center}
  \begin{tabular}{lllll}
    \hline
    $\phi(1020)$ &
    $h_1(1170)$ &
    $b_1(1235)$ &
    $a_1(1260)$ &
    $f_2(1270)$ \\
    $f_1(1285)$ &
    $\eta(1295)$ &
    $\pi(1300)$ &
    $a_2(1320)$ & 
    $f_0(1370)$ \\
    $\pi_1(1400)$ &
    $\eta(1405)$ &
    $h_1(1415)$ &
    $f_1(1420)$ &
    $\omega(1420)$ \\
    $a_0(1450)$ &
    $\rho(1450)$ &
    $\eta(1475)$ &
    $f_0(1500)$ & $\ldots$\\
    \hline
  \end{tabular}
\end{center}
\caption{A list of resonances between 1 and 1.5 GeV.}
\label{tab:resonances}
\end{table}

Let's be a bit clearer about short-distance constraints, there are different types:
\begin{itemize}[noitemsep,nolistsep]
  \item Those used for constraints on hadronic form-factors
  \begin{itemize}[noitemsep,nolistsep]
    \item Couplings of hadrons to off-shell photons
    \begin{itemize}[noitemsep,nolistsep]
      \item Pure OPE (e.g. $\pi^0\to\gamma^*\gamma^*$ at $Q_1^2=Q_2^2$)
      \item Brodsky-Lepage-Radyushkin-$\cdots$:
  \item the overall power is very well predicted (counting rules)
  \item the coefficient follows from the asymptotic wave functions and possible $\alpha_S$ corrections: larger uncertainty
  \end{itemize}
  \item Light-cone QCD sum rules
  \item This type is mainly used in HLbL to put constraints on the form-factors
      in the individual contributions and we do not discuss them more. 
  \end{itemize}
  \item Those putting constraints on the full four-point function (4, 3 or 2 currents close)
  \begin{itemize}[noitemsep,nolistsep]
  \item {SD4:} $\Pi^{\mu\nu\lambda\sigma} (q_1,q_2,q_3)$ $Q_1,Q_2,Q_3$ and $Q_4$ large with all $Q_i\cdot Q_j$ large:  the standard OPE
\item {SD3:} $\left. \dfrac{\delta \Pi^{\mu\nu\lambda\sigma}(q_1,q_2,q_3)}{ \delta  q_{4\rho}}\right|_{q_4=0}$ with
$Q_1^2\sim Q_2^2 \sim Q_3^2 \gg\Lambda_{QCD}^2$ and $Q_4=0$ \cite{Bijnens:2019ghy,Bijnens:2020xnl,Bijnens:2021jqo}
\item {SD2:} $\left. \dfrac{\delta \Pi^{\mu\nu\lambda\sigma}(q_1,q_2,q_3)}{ \delta  q_{4\rho}}\right|_{q_4=0}$\hskip-2mm and $Q_1^2\sim Q_2^2 \gg Q_3^2,\Lambda_{QCD}^2$ and $Q_4=0$ \cite{Melnikov:2003xd,Bijnens:2022itw,Bijnens:2024jgh}
\end{itemize}
\end{itemize}

\section{SD4 and SD3}

\begin{figure}
  \begin{center}
    \raisebox{1.5cm}{
      \setlength\unitlength{1.4pt}
      \begin{axopicture}(60,70)
      \SetScale{1.4}  
      \Photon(0,0)(15,10){2.5}{3.5}
      \ArrowLine(15,10)(35,10)
      \ArrowLine(35,10)(55,10)
      \Text(35,10){$\otimes$}
      \Text(35,5)[t]{``$q_4$''}
      \ArrowLine(55,10)(35,40)
      \ArrowLine(35,40)(15,10)
      \Photon(55,10)(70,0){2.5}{3.5}
      \Photon(35,40)(35,55){2.5}{3.5}
      \Text(5,10)[r]{$q_1$}
      \Text(65,10)[l]{$q_2$}
      \Text(38,48)[l]{$q_3$}
      \Arc[arrow](35,22)(7,-80,260)
      \Text(35,21){$p$}
      \end{axopicture}}
      \hskip2cm
        \includegraphics[width=8cm]{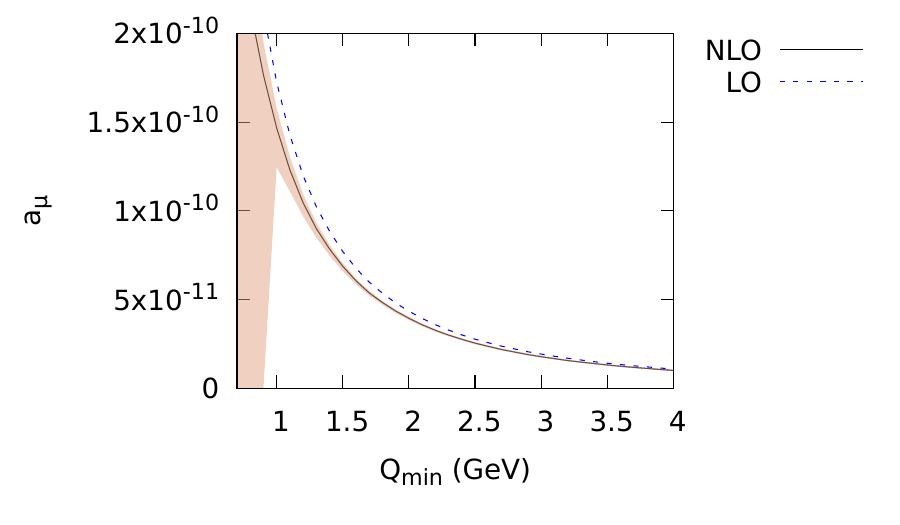}
        \vspace*{-1.cm}  
  \end{center}
  \caption{Left: The leading term in the background OPE, the cross is the background insertion. Right: The contribution from the massless quark-loop and gluonic corrections, both lowest order in the SD3 OPE as a function of the lower cut-off $Q_\text{min}$ for $Q_1,Q_2,Q_3$. The error is from the running of $\alpha_S$ and explodes for a low cut-off. Figure from \cite{Bijnens:2021jqo}.}
  \label{fig:gluonic}
\end{figure}

The SD4 type might be useful for putting constraints on modelling the four-point function but it is not clear how relevant it can be since it requires $Q_4$ large and we need results in the limit $Q_4\to0$. We really need the case SD3. We realized in \cite{Bijnens:2019ghy} that a similar problem had been solved earlier in the context of QCD sum rules for electromagnetic form-factors \cite{Balitsky:1983xk,Ioffe:1983ju}\footnote{We later realized that \cite{Czarnecki:2002nt} had used the same method for the electroweak two-loop contribution.}. The essence is that for the $q_4$ leg in \cref{fig:hadronic}(b,c) we use a constant background field and we do the OPE of the three other currents, the $q_1,q_2,q_3$ legs, in the presence of that background field. Technically this is done using the radial gauge $A_4^\lambda(w) = \frac{1}{2}w_\mu F^{\mu\lambda}$ and we will need condensates in the presence of the background field.
The first term is exactly the massless quark loop, shown schematically in \cref{fig:gluonic} and the next term combines perturbative quark-mass corrections with the vacuum susceptibility, only the sum of the two is well defined \cite{Bijnens:2019ghy}. We have since evaluated the next-order in the OPE (NNLO) \cite{Bijnens:2020xnl} as well as the gluonic corrections to the massless quark-loop \cite{Bijnens:2021jqo}. For a lower cut-off above 1.4 GeV or so both the susceptibility, quark-mass and the NNLO OPE do not contribute at the precision needed. The gluonic corrections are of order $-10\%$ of the massless quark-loop and are shown in \cref{fig:gluonic}. Both the quark-loop contribution and the gluonic correction to the derivative of the four-point function \cref{eq:derivpi} and the $\overline{\Pi}_i$ of \cref{eq:amuint} are known fully analytically in terms of polylogarithms.

\section{SD2, corner or Melnikov-Vainshtein short-distance}
\subsection{OPE}

\begin{figure}
  \begin{center}
   \begin{minipage}{6cm} 
   \begin{overpic}[percent,width=0.99\textwidth]{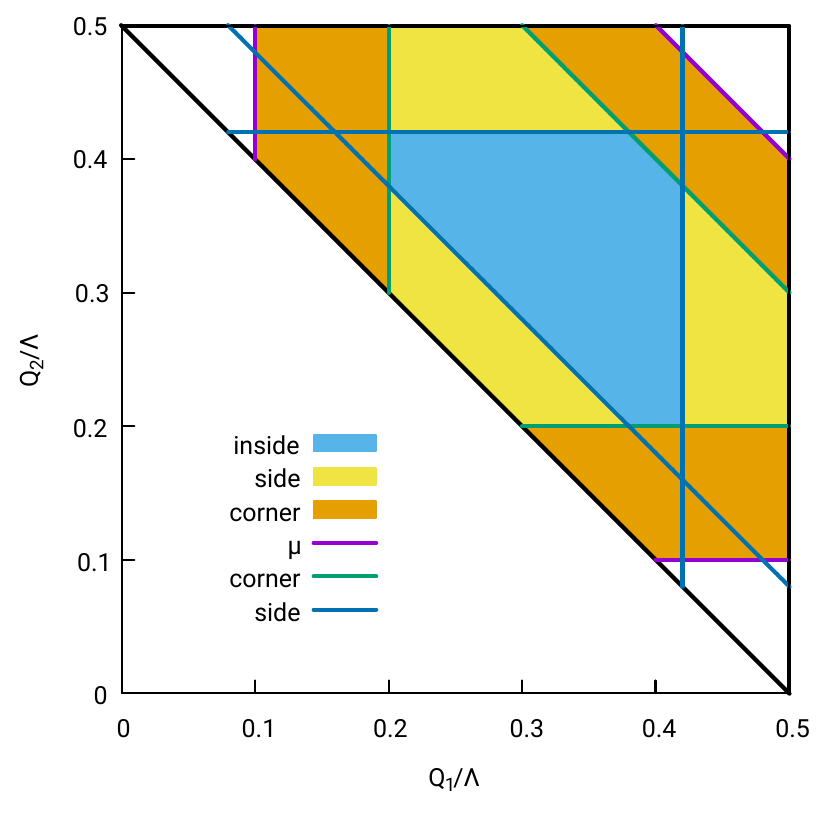}
      \put(90,100){$Q_3$ small}
      \put(10,100){$Q_1$ small}
      \put(90,5){$Q_2$ small}
    \end{overpic}
      \vskip-0.5cm
   \centerline{(a)}
  \end{minipage}
  \hspace*{2cm}
\raisebox{-1cm}{
  \setlength{\unitlength}{1pt}
    \begin{axopicture}(100,40)(0,-15)
    \SetScale{1.0}
    \ArrowLine(0,3)(10,3)
    \ArrowLine(10,3)(30,3)
    \ArrowLine(30,3)(40,3)
    \Photon(12,3)(12,40){2}{6}
    \Photon(28,3)(28,40){2}{6}
    \Text(50,3)[b]{+}
    \ArrowLine(60,3)(70,3)
    \ArrowLine(70,3)(90,3)
    \ArrowLine(90,3)(100,3)
    \Photon(72,3)(88,40){2}{8}
    \Photon(88,3)(72,40){2}{8}
    \Text(50,-15)[b]{(b)}
    \end{axopicture}}
    \end{center}
  \caption{(a) The different integration regions for fixed $\Lambda=Q_1+Q_2+Q_3$ (b) The LO diagrams for the expansion of two-currents close to each other. The leading term gives an axial current.}
  \label{fig:expansion}
  \end{figure}

This is in the regime called SD2 above. We also call it the corner regime because if we look at the integration region of  $Q_1,Q_2,Q_3$ for a fixed $\Lambda=Q_1+Q_2+Q_3$ as shown in \cref{fig:expansion}. The integration region is the large triangle in the upper right. The SD2 regions are the three corners. The orange can be treated in perturbative QCD, the white parts need nonperturbative input.
Crosschecks are possible in the regions where both SD2 and SD3 are applicable.
  
It was first studied in \cite{Melnikov:2003xd} and corresponds to the case where one of the $Q_i$ is much smaller than the other two (at $Q_4=0$), say $Q_3$. This implies from $q_1+q_2+q_3=0$ that $Q_1^2\approx Q_2^2\gg Q_3^q,(Q_1-Q_2)^2$. Expanding the two electromagnetic currents close to each other shown schematically in \cref{fig:expansion}(a) gives an axial current and thus \cite{Melnikov:2003xd}
\begin{align}
  \label{eq:MV}
  \Pi^{\rho\nu\alpha\beta}\propto \frac{P^\rho}{Q_1^2}\langle 0|
  T\left(J_A^\nu J_V^\alpha J_V^\beta\right)|0\rangle
\end{align}
The matrix element in \cref{eq:MV} corresponds to the anomaly and a lot is thus known about it both perturbatively and nonperturbatively, in particular nonrenormalization theorems apply. This has been studied extensively since and is discussed in detail in \cite{Aoyama:2020ynm}. More recent references are in \cite{Ludtke:2020moa}.

In the chiral limit, where \cref{eq:MV} is valid, there is only a prediction for one of the five needed intermediate functions. This is \cite{Colangelo:2021nkr}
\begin{align}
  \label{eq:MV2}
    \hat\Pi_1 = \dfrac{e_q^4}{\pi^2}\dfrac{-12}{Q_3^2\overline Q_3^2}\left(1-\dfrac{\alpha_S}{\pi}\right),
    \qquad \overline{Q}_3\equiv Q_1+Q_2,\qquad Q_3\ll Q_1,Q_2\,,
\end{align}
where the $\alpha_S$ correction was conjectured in \cite{Ludtke:2020moa}. In the SD3 regime our calculations \cite{Bijnens:2021jqo,Bijnens:2021jqo} confirmed \cref{eq:MV2}. We have since in \cite{Bijnens:2024jgh} also confirmed it fully.
The next term in the OPE we could look at by expanding the SD3 results from \cite{Bijnens:2001cq} and already had two distinctive features. $\log\left(Q_3^2/\overline{Q}_3^2\right)$ already shows up at $\alpha_S=0$ and for some of the other $\hat\Pi_i$ the gluonic terms dominate.

To start we define\footnote{The factor $1/e^2$ cancels the $e$ factors that occur due to the matrix element with two photons.}
\begin{align}
\label{eq:twopoint}
\Pi^{\mu_{1}\mu_{2}}
=
\, 
\frac{i}{e^{2}}\int d^{4}x_1\int d^{4}x_2 e^{-i(q_1 x_1+q_2 x_2)}\langle 0 |T(J^{\mu_1}(x_1)J^{\mu_2}(x_2))|\gamma(q_3)\gamma(q_4) \rangle 
\end{align}
This object allows to get the four point function needed for $a_\mu$ via
$
\Pi^{\mu_1\mu_2}=\epsilon_{\mu_3}\epsilon_{\nu_4}\Pi^{\mu_1\mu_2\mu_3\nu_4} \, ,
$
and the needed $\partial/\partial q_{4,\mu_4}$ at $q_4\to0$ as well.
To obtain the OPE on the two currents we need to work out
\begin{align}
\int d^{4}x_1\int d^{4}x_2 e^{-i(q_1 x_1+q_2 x_2)}T(J^{\mu_1}(x_1)J^{\mu_2}(x_2))
\end{align}
for
$\hat q = (q_1-q_2)/2$ with $\hat Q^2=-\hat q$ large, $q_3=-q_1-q_2$ is small.
\begin{figure}
  \begin{center}
    \begin{minipage}{3cm}
    \includegraphics[width=0.99\textwidth]{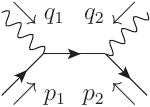}
    \centerline{(a)}
    \end{minipage}
    \hspace*{2cm}
    \raisebox{-1.3cm}{
    \begin{axopicture}(70,85)(0,-15)
      \Photon(0,0)(20,20){2.5}{3.5}
      \Photon(0,70)(20,50){2.5}{3.5}
      \Photon(70,70)(50,50){2.5}{3.5}
      \Photon(70,0)(50,20){2.5}{3.5}
      \ArrowLine(20,20)(50,20)
      \ArrowLine(50,20)(50,50)
      \ArrowLine(50,50)(20,50)
      \ArrowLine(20,50)(20,20)
      \Text(60,15)[l]{$q_4$}
      \Text(10,15)[r]{$q_1$}
      \Text(10,55)[r]{$q_2$}
      \Text(60,55)[l]{$q_3$}
      \Text(35,-15)[b]{(b)}
      %\Arc[arrow](35,35)(9,-80,260)
      %\Text(35,35){$p$}
      \end{axopicture}}
      \hspace*{2cm}
      \raisebox{-1.3cm}{
      \begin{axopicture}(70,75)(0,-15)
        \Photon(0,0)(15,10){2.5}{3.5}
        \ArrowLine(15,10)(55,10)
        \ArrowLine(55,10)(35,40)
        \ArrowLine(35,40)(15,10)
        \Photon(55,10)(70,0){2.5}{3.5}
        %\Photon(35,40)(35,55){2.5}{3.5}
        \Text(35,40){$\otimes$}
        \Text(5,10)[r]{$q_3$}
        \Text(65,10)[l]{$q_4$}
        \Text(35,48)[c]{$O_i$}
        \Text(35,-15)[b]{(c)}
        %\Arc[arrow](35,22)(7,-80,260)
        %\Text(35,21){$p$}
        \end{axopicture}}
  \end{center}
  \caption{(a) The diagram to be expanded in $\hat q=q_1-q_2$ large together with the permuation $q_1\leftrightarrow q_2$ (b) The type of diagrams that leads to extra field strength operators} 
  \label{fig:OPE}
\end{figure}
So we work out the expansion from the diagram in \cref{fig:OPE}(a) and its permutation
\begin{align}
  \label{eq:expansionquark}
  %\nonumber
  %
  &\Pi^{\mu_{1}\mu_{2}}_{\bar{q}q}  
  \approx - \frac{ e_q^2}{e^2}\, \frac{-\hat q_\alpha}{\hat q^2}  \langle0|\bar{q}(0)[\gamma^{\mu_1}\gamma^\alpha\gamma^{\mu_2}-\gamma^{\mu_2}\gamma^\alpha\gamma^{\mu_1}]q(0))|\gamma(q_3)\gamma(q_4) \rangle
  \\&
  -\frac{ie_q^2}{e^2\hat{q}^2}(g_{\mu_1\delta}g_{\mu_2\beta}+g_{\mu_2\delta}g_{\mu_1\beta}-g_{\mu_1\mu_2}g_{\delta\beta})
  %\nonumber\\&\hskip2cm\times
  \left(g^{\alpha\delta}-2\frac{\hat{q}^\delta \hat{q}^\alpha}{\hat{q}^2}\right) \langle0| \bar{q}(0)(\overrightarrow{D}^{\alpha}-\overleftarrow{D}^{\alpha})\gamma^{\beta}  q(0))|\gamma(q_3)\gamma(q_4) \rangle  
  \nonumber
  \end{align}
The first line is the $D=3$ contribution studied by \cite{Melnikov:2003xd} and many others. The second in \cref{eq:expansionquark} is the $D=4$ operators. However, this is not the full story, the type of diagrams in \cref{fig:OPE}(b) leads to operators with only field strengths, i.e. combinations of  $F_{\mu\nu}F_{\alpha\beta}$ and $G_{\mu\nu}G_{\alpha\beta}$ that also contribute to the matrix element in \cref{eq:twopoint}. We now need to add gluonic corrections as well and use the equations of motion to obtain the operators that can contribute, these are given as Eqs. (3.7) to (3.15) in \cite{Bijnens:2024jgh}.

The relevant Wilson coefficients have been calculated at the high scale and all dependence on\footnote{And of course the relevant permutations for the other corners.} $\hat q$ is in the Wilson coefficients. The matrix elements only depend on $q_3$. We can use standard RGE methods to run the Wilson coefficients down to a lower scale. A number of comments are in order:
\begin{itemize}[noitemsep,nolistsep]
  \item  We now have $\lim_{q_4\to0}(\partial/\partial q_{4,\mu_4})\Pi^{\mu_1\mu_2\mu_3\nu_4}$ to two powers in $\hat q$
  \item Note gauge invariance for $q_3$ is exact
  \item $q_4\to0$ gauge invariance is antisymmetry in $\nu_4\mu_4$
  \item BUT gauge invariance for $q_1,q_2$ only perturbatively in $\hat q$
  \item{Consequence: be careful when using gauge equivalent expressions}
  \item In particular when using projectors to get quantities without Lorentz indices (our intermediate $\widetilde\Pi$ or the $\hat\Pi_i$) {need to use the projectors with lowest powers of $\hat q$ possible}   
  \end{itemize}
  
\subsection{Perturbative matrix elements}

The matrix elements of the quark operators can be calculated using diagrams as those in \cref{fig:OPE}(c) and gluonic corrections. The $D=3$ part of \cref{eq:expansionquark} reproduces \cref{eq:MV2} and that there is no contribution from this order in the other corners.

The $D=4$ matrix elements are more subtle, as an example we obtain
\begin{align}
\label{eq:singularity1}  
\hat\Pi_7(O_1)=\frac{32\delta_{12} Q_3^2}{3\pi^2(Q_3^2-\delta_{12}^2)\overline Q_3^5}
\text{ for }Q_3\text{ and }\delta_{12}=Q_1-Q_2\text{ small}\,,
\end{align}
which has a kinematic singularity for $\delta_{12}=\pm Q_3$. Similar singularities occur in other functions. However, when relating the matrix elements of the operators to the $\hat\Pi_i$ one needs to be careful with keeping powers of $\hat q$ also in the needed projectors which leads to mixing of the $D=3$ and $D=4$ operators at the same $\hat q$ order in the $\hat\Pi_i$. The perturbative matrix elements are such that when combined with the Wilson coefficients for the various operators into the $\hat\Pi_i$ all these kinematic singularities cancel.

As an example we give the results in the corner for $Q_2$ small without gluonic corrections
   \begin{align}
    \hat{\Pi}_{1} & =   -\frac{16 \left( 5+6\, \log 2\, \frac{Q_2}{\overline{Q}_{2}}\right)  }{9\pi^2 \,   \overline{Q}_{2}^4} 
     +\mathcal{O}\left( \overline{Q}_{2}^{-5}\right) &
%     \nonumber\\
     \hat{\Pi}_{4} & = -\frac{4}{3\pi^2 \, Q_{2}^2  \, \overline{Q}_{2}^2}
      +\mathcal{O}\left(\overline{Q}_{2}^{-3}\right)
         \nonumber\\
     \hat{\Pi}_{7} & = -\frac{16}{3 \pi^2 \, Q_2^2 \, \overline{Q}_2^4}
      +\mathcal{O}\left(\overline{Q}_{2}^{-5}\right) &
%         \nonumber\\
      \hat{\Pi}_{17} & = 
      \mathcal{O}\left( \overline{Q}_{2}^{-5}\right)
         \nonumber\\
     \hat{\Pi}_{39} & = \frac{16}{3\pi^2 \, Q_2^2 \,  \overline{Q}_{2}^4} 
      +\mathcal{O}\left( \overline{Q}_{2}^{-5}\right) &
 %        \nonumber\\
     \hat{\Pi}_{54} & = -\frac{8}{3\pi^2 \, Q_2^2 \, \overline{Q}_2^4}
      +\mathcal{O}\left( \overline{Q}_{2}^{-5}\right)\,.
   \end{align}
$\overline Q_2=Q_3+Q_1$. The results for the other corners are in \cite{Bijnens:2022itw} and with gluonic corrections in \cite{Bijnens:2024jgh}.   

\subsection{Nonperturbative matrix elements}

The $D=3$ matrix element leads to the general expression \cite{Melnikov:2003xd} for general $q_3^2$, also small compared to the QCD scale:
\begin{align}
  \label{eq:D3}
  \lim_{q_4 \rightarrow 0} & \frac{\partial\Pi^{\mu_1\mu_2\mu_3  \nu_4}}{\partial q_{4,\, \mu_4}}
  =\frac{-q_{3}^2 \epsilon^{\mu_1\mu_2\hat{q}\delta}}{2\pi^2\hat{q}^2}
  \Big(\epsilon_{\mu_3\mu_4\nu_4\delta}\, \omega_{T}(q_3^{2})-\frac{q_{3\mu_3}}{q_3^2}\, \epsilon_{q_3\mu_4\nu_4\delta}\,\, \omega_{T}(q_3^{2})
  %\nonumber \\
  %&
%  \hspace{27.5ex}
  +\frac{q_{3\delta}}{q_3^2}\, \epsilon_{\mu_3 \mu_4 \nu_4 q_3}\, \, \left[ \omega_{L}(q_3^2)-\omega_{T}(q_3^2)\right]\Big) \, .
  \nonumber\end{align}
This fully agrees with \cite{Melnikov:2003xd} and the later discussions giving in the chiral limit
 \begin{align}
 \hat{\Pi}_{1}& =\frac{2}{\pi^2\overline{Q}_{3}^2}\, \omega_{L}(q_3^2)\,.
 \end{align}
The contribution from \cref{eq:D3} to the $\hat\Pi_i$ at the next order in $1/\hat q$ is (for  the $q_2$ small corner and with $\delta_{31}=Q_3-Q_1$)
\begin{align}
  \label{singularity2}
  \hat{\Pi}_{4}&=\frac{ Q_2^2}{\pi^2(Q_2^2-\delta_{31}^2)\overline{Q}_2^2}\, \omega_T(q_2^2)&
  %\nonumber\\
  \hat{\Pi}_{7}&=\frac{4 Q_2^2}{\pi^2(Q_2^2-\delta_{31}^2)\overline{Q}_2^4}\, \omega_T(q_2^2)
  \\
  \hat{\Pi}_{39}&=-\frac{4 Q_2^2}{\pi^2(Q_2^2-\delta_{31}^2)\overline{Q}_2^4}\, \omega_T(q_2^2)&
  %\nonumber\\
  \hat{\Pi}_{54}&=\frac{4 Q_2^2}{\pi^2(Q_2^2-\delta_{31}^2)\overline{Q}_2^4}\, \omega_T(q_2^2)\,.
  \end{align}
This has the same type of singularity as in \cref{eq:singularity1} and they must cancel against similar singularities coming from the matrix elements of the $D=4$ operators that contribute to the same order in $1/\hat q$. This requirement leads  to three relations between the form-factors of the $D=4$ operators and $\omega_T$
\begin{align}
  \label{eq:relationD3D4}
      \omega _{(8)}^{D,2} & = -2\, \omega _{(8)}^{D,1}+ \omega _{(8)}^{D,5} -\frac{\omega _{(8)}^{D,6}}{2}-\frac{\omega _{T,(8)}Q_i^2}{8\pi^2 } \, ,&
      \omega _{(8)}^{D,4} & =\omega _{(8)}^{D,5},
      \nonumber \\
      \omega _{(8)}^{D,3} & = -2\, \omega _{(8)}^{D,1}+ \omega _{(8)}^{D,5} -\frac{\omega _{(8)}^{D,6}}{2}+\frac{\omega _{T,(8)}Q_i^2}{8\pi^2 }
  \end{align}

  At the time of the talk this were the results we had. Since then we have realized a rather large cancellation in the contribution to $a_\mu$. In \cref{eq:amuint} we also expand the $T_i$ in powers of $1/\hat q$ and then the combined contribution from the $D=3$ and $D=4$ operators to $a_\mu$ after using the relations of \cref{eq:relationD3D4} can be fully written in terms of $\omega_L$ and $\omega_T$.
  This is discussed in more detail in \cite{Bijnens:2024jgh} and has the  consequence that to NLO in $1/\hat q$ no new phenomenological information is needed greatly simplifying the next order in the corner expansion.

\section{Conclusions}

We have briefly discussed our work on the short-distance calculation with QCD of the HLbL contribution to the muon anomalous magnetic moment. We expect the results described here and in \cite{Bijnens:2019ghy,Bijnens:2020xnl,Bijnens:2021jqo,Bijnens:2022itw,Bijnens:2024jgh} to play a significant role in the reduction of the error on the HLbL contribution.

\section*{Acknowledgements}
N.~H.-T.~is funded by the UK Research and Innovation, Engineering and Physical Sciences Research Council, grant number EP/X021971/1.
% N.~H.-T.~wishes to thank Lund University for hosting him while parts of the project were completed. 
A.~R.-S.~is funded in part by MIUR contract number 2017L5W2PT and by the Generalitat Valenciana (Spain) through the plan GenT program (CIDEIG/2023/12) and by the Spanish Government (Agencia Estatal de Investigación MCIN/AEI/\-10.13039/501100011033) Grant No. PID2023-146220NB-I00. 

\bibliographystyle{JHEPsmall}
\bibliography{refs}

\providecommand{\href}[2]{#2}\begingroup\raggedright\begin{thebibliography}{10}\setlength\itemsep{0cm}

\bibitem{Bennett:2006fi}
{\scshape Muon g-2} collaboration, \emph{{Final Report of the Muon E821
  Anomalous Magnetic Moment Measurement at BNL}},
  \href{https://doi.org/10.1103/PhysRevD.73.072003}{\emph{Phys. Rev.}
  {\bfseries D73} (2006) 072003}
  [\href{https://arxiv.org/abs/hep-ex/0602035}{{\ttfamily hep-ex/0602035}}].

\bibitem{Muong-2:2023cdq}
{\scshape Muon g-2} collaboration, \emph{{Measurement of the Positive Muon
  Anomalous Magnetic Moment to 0.20~ppm}},
  \href{https://doi.org/10.1103/PhysRevLett.131.161802}{\emph{Phys. Rev. Lett.}
  {\bfseries 131} (2023) 161802}
  [\href{https://arxiv.org/abs/2308.06230}{{\ttfamily 2308.06230}}].

\bibitem{Aoyama:2020ynm}
T.~Aoyama et~al., \emph{{The anomalous magnetic moment of the muon in the
  Standard Model}},
  \href{https://doi.org/10.1016/j.physrep.2020.07.006}{\emph{Phys. Rept.}
  {\bfseries 887} (2020) 1} [\href{https://arxiv.org/abs/2006.04822}{{\ttfamily
  2006.04822}}].

\bibitem{ColangeloPoS}
G.~Colangelo, \emph{{Dispersive approach to hadronic contributions to the Muon
  $g-2$}}, {\emph{PoS} {\bfseries CD2024} (2025) 016}.

\bibitem{WittigPoS}
H.~Wittig, \emph{{The puzzles of the muon anomalous magnetic moment}},
  {\emph{PoS} {\bfseries CD2024} (2025) 017}.

\bibitem{HolzPoS}
S.~Holz, \emph{{Dispersive determination of the $\eta/\eta'$ transition form
  factors}}, {\emph{PoS} {\bfseries CD2024} (2025) 036}.

\bibitem{Hayakawa:1997rq}
M.~Hayakawa and T.~Kinoshita, \emph{{Pseudoscalar pole terms in the hadronic
  light by light scattering contribution to muon g - 2}},
  \href{https://doi.org/10.1103/PhysRevD.57.465}{\emph{Phys. Rev. D} {\bfseries
  57} (1998) 465} [\href{https://arxiv.org/abs/hep-ph/9708227}{{\ttfamily
  hep-ph/9708227}}].

\bibitem{Hayakawa:2001bb}
M.~Hayakawa and T.~Kinoshita, \emph{{Comment on the sign of the pseudoscalar
  pole contribution to the muon g-2}},
  \href{https://arxiv.org/abs/hep-ph/0112102}{{\ttfamily hep-ph/0112102}}.

\bibitem{Bijnens:1995cc}
J.~Bijnens, E.~Pallante and J.~Prades, \emph{{Hadronic light by light
  contributions to the muon g-2 in the large N(c) limit}},
  \href{https://doi.org/10.1103/PhysRevLett.75.1447}{\emph{Phys. Rev. Lett.}
  {\bfseries 75} (1995) 1447}
  [\href{https://arxiv.org/abs/hep-ph/9505251}{{\ttfamily hep-ph/9505251}}].

\bibitem{Bijnens:1995xf}
J.~Bijnens, E.~Pallante and J.~Prades, \emph{{Analysis of the hadronic light by
  light contributions to the muon g-2}},
  \href{https://doi.org/10.1016/0550-3213(96)00288-X}{\emph{Nucl. Phys. B}
  {\bfseries 474} (1996) 379}
  [\href{https://arxiv.org/abs/hep-ph/9511388}{{\ttfamily hep-ph/9511388}}].

\bibitem{Bijnens:2001cq}
J.~Bijnens, E.~Pallante and J.~Prades, \emph{{Comment on the pion pole part of
  the light by light contribution to the muon g-2}},
  \href{https://doi.org/10.1016/S0550-3213(02)00074-3}{\emph{Nucl. Phys. B}
  {\bfseries 626} (2002) 410}
  [\href{https://arxiv.org/abs/hep-ph/0112255}{{\ttfamily hep-ph/0112255}}].

\bibitem{Knecht:2001qf}
M.~Knecht and A.~Nyffeler, \emph{{Hadronic light by light corrections to the
  muon g-2: The Pion pole contribution}},
  \href{https://doi.org/10.1103/PhysRevD.65.073034}{\emph{Phys. Rev. D}
  {\bfseries 65} (2002) 073034}
  [\href{https://arxiv.org/abs/hep-ph/0111058}{{\ttfamily hep-ph/0111058}}].

\bibitem{Melnikov:2003xd}
K.~Melnikov and A.~Vainshtein, \emph{{Hadronic light-by-light scattering
  contribution to the muon anomalous magnetic moment revisited}},
  \href{https://doi.org/10.1103/PhysRevD.70.113006}{\emph{Phys. Rev.}
  {\bfseries D70} (2004) 113006}
  [\href{https://arxiv.org/abs/hep-ph/0312226}{{\ttfamily hep-ph/0312226}}].

\bibitem{Aldins:1970id}
J.~Aldins, T.~Kinoshita, S.J.~Brodsky and A.J.~Dufner, \emph{Photon - photon
  scattering contribution to the sixth order magnetic moments of the muon and
  electron}, \href{https://doi.org/10.1103/PhysRevD.1.2378}{\emph{Phys. Rev.}
  {\bfseries D1} (1970) 2378}.

\bibitem{Colangelo:2015ama}
G.~Colangelo, M.~Hoferichter, M.~Procura and P.~Stoffer, \emph{{Dispersion
  relation for hadronic light-by-light scattering: theoretical foundations}},
  \href{https://doi.org/10.1007/JHEP09(2015)074}{\emph{JHEP} {\bfseries 09}
  (2015) 074} [\href{https://arxiv.org/abs/1506.01386}{{\ttfamily
  1506.01386}}].

\bibitem{Bijnens:2019ghy}
J.~Bijnens, N.~Hermansson-Truedsson and A.~Rodr{\'i}guez-S{\'a}nchez,
  \emph{{Short-distance constraints for the HLbL contribution to the muon
  anomalous magnetic moment}},
  \href{https://doi.org/10.1016/j.physletb.2019.134994}{\emph{Phys. Lett.}
  {\bfseries B798} (2019) 134994}
  [\href{https://arxiv.org/abs/1908.03331}{{\ttfamily 1908.03331}}].

\bibitem{Blum:2023vlm}
{\scshape RBC, UKQCD} collaboration, \emph{{Hadronic light-by-light
  contribution to the muon anomaly from lattice QCD with infinite volume QED at
  physical pion mass}},
  \href{https://doi.org/10.1103/PhysRevD.111.014501}{\emph{Phys. Rev. D}
  {\bfseries 111} (2025) 014501}
  [\href{https://arxiv.org/abs/2304.04423}{{\ttfamily 2304.04423}}].

\bibitem{Chao:2021tvp}
E.-H.~Chao, R.J.~Hudspith, A.~G\'erardin, J.R.~Green, H.B.~Meyer and K.~Ottnad,
  \emph{{Hadronic light-by-light contribution to $(g-2)_\mu $ from lattice QCD:
  a complete calculation}},
  \href{https://doi.org/10.1140/epjc/s10052-021-09455-4}{\emph{Eur. Phys. J. C}
  {\bfseries 81} (2021) 651}
  [\href{https://arxiv.org/abs/2104.02632}{{\ttfamily 2104.02632}}].

\bibitem{Fodor:2024jyn}
Z.~Fodor, A.~Gerardin, L.~Lellouch, K.K.~Szabo, B.C.~Toth and C.~Zimmermann,
  \emph{{Hadronic light-by-light scattering contribution to the anomalous
  magnetic moment of the muon at the physical pion mass}},
  \href{https://arxiv.org/abs/2411.11719}{{\ttfamily 2411.11719}}.

\bibitem{Hoferichter:2024bae}
M.~Hoferichter, P.~Stoffer and M.~Zillinger, \emph{{Dispersion relation for
  hadronic light-by-light scattering: subleading contributions}},
  \href{https://doi.org/10.1007/JHEP02(2025)121}{\emph{JHEP} {\bfseries 02}
  (2025) 121} [\href{https://arxiv.org/abs/2412.00178}{{\ttfamily
  2412.00178}}].

\bibitem{Mager:2025pvz}
J.~Mager, L.~Cappiello, J.~Leutgeb and A.~Rebhan, \emph{{Longitudinal
  short-distance constraints on hadronic light-by-light scattering and tensor
  meson contributions to the muon $g-2$}},
  \href{https://arxiv.org/abs/2501.19293}{{\ttfamily 2501.19293}}.

\bibitem{Bijnens:2020xnl}
J.~Bijnens, N.~Hermansson-Truedsson, L.~Laub and A.~Rodr\'\i{}guez-S\'anchez,
  \emph{{Short-distance HLbL contributions to the muon anomalous magnetic
  moment beyond perturbation theory}},
  \href{https://doi.org/10.1007/JHEP10(2020)203}{\emph{JHEP} {\bfseries 10}
  (2020) 203} [\href{https://arxiv.org/abs/2008.13487}{{\ttfamily
  2008.13487}}].

\bibitem{Bijnens:2021jqo}
J.~Bijnens, N.~Hermansson-Truedsson, L.~Laub and A.~Rodr\'\i{}guez-S\'anchez,
  \emph{{The two-loop perturbative correction to the $(g - 2)_\mu$ HLbL at
  short distances}}, \href{https://doi.org/10.1007/JHEP04(2021)240}{\emph{JHEP}
  {\bfseries 04} (2021) 240}
  [\href{https://arxiv.org/abs/2101.09169}{{\ttfamily 2101.09169}}].

\bibitem{Bijnens:2022itw}
J.~Bijnens, N.~Hermansson-Truedsson and A.~Rodr\'\i{}guez-S\'anchez,
  \emph{{Constraints on the hadronic light-by-light in the Melnikov-Vainshtein
  regime}}, \href{https://doi.org/10.1007/JHEP02(2023)167}{\emph{JHEP}
  {\bfseries 02} (2023) 167}
  [\href{https://arxiv.org/abs/2211.17183}{{\ttfamily 2211.17183}}].

\bibitem{Bijnens:2024jgh}
J.~Bijnens, N.~Hermansson-Truedsson and A.~Rodr\'\i{}guez-S\'anchez,
  \emph{{Constraints on the hadronic light-by-light tensor in corner kinematics
  for the muon $g-2$}},  \href{https://arxiv.org/abs/2411.09578}{{\ttfamily
  2411.09578}}.

\bibitem{Balitsky:1983xk}
I.I.~Balitsky and A.V.~Yung, \emph{{Proton and Neutron Magnetic Moments from
  QCD Sum Rules}},
  \href{https://doi.org/10.1016/0370-2693(83)90676-7}{\emph{Phys. Lett.}
  {\bfseries 129B} (1983) 328}.

\bibitem{Ioffe:1983ju}
B.L.~Ioffe and A.V.~Smilga, \emph{{Nucleon Magnetic Moments and Magnetic
  Properties of Vacuum in QCD}},
  \href{https://doi.org/10.1016/0550-3213(84)90364-X}{\emph{Nucl. Phys.}
  {\bfseries B232} (1984) 109}.

\bibitem{Czarnecki:2002nt}
A.~Czarnecki, W.J.~Marciano and A.~Vainshtein, \emph{{Refinements in
  electroweak contributions to the muon anomalous magnetic moment}},
  \href{https://doi.org/10.1103/PhysRevD.67.073006}{\emph{Phys. Rev.}
  {\bfseries D67} (2003) 073006}
  [\href{https://arxiv.org/abs/hep-ph/0212229}{{\ttfamily hep-ph/0212229}}].

\bibitem{Ludtke:2020moa}
J.~L\"udtke and M.~Procura, \emph{{Effects of Longitudinal Short-Distance
  Constraints on the Hadronic Light-by-Light Contribution to the Muon $g-2$}},
  \href{https://doi.org/10.1140/epjc/s10052-020-08611-6}{\emph{Eur. Phys. J. C}
  {\bfseries 80} (2020) 1108}
  [\href{https://arxiv.org/abs/2006.00007}{{\ttfamily 2006.00007}}].

\bibitem{Colangelo:2021nkr}
G.~Colangelo, F.~Hagelstein, M.~Hoferichter, L.~Laub and P.~Stoffer,
  \emph{{Short-distance constraints for the longitudinal component of the
  hadronic light-by-light amplitude: an update}},
  \href{https://doi.org/10.1140/epjc/s10052-021-09513-x}{\emph{Eur. Phys. J. C}
  {\bfseries 81} (2021) 702}
  [\href{https://arxiv.org/abs/2106.13222}{{\ttfamily 2106.13222}}].

\end{thebibliography}\endgroup

%\begin{thebibliography}{99}
%\bibitem{...}

%\end{thebibliography}

\end{document}